\documentclass[12pt]{iopart}
\usepackage{iopams}
\usepackage{setstack}

\newcommand{\half}{\textstyle{\frac{1}{2}}}

\newcommand{\cP}{{\cal P}}
\newcommand{\cT}{{\cal T}}

\newcommand{\tsdp}{\textstyle{\frac{\rm d}{{\rm d}p}}}
\newcommand{\tsdt}{\textstyle{\frac{\rm d}{{\rm d}t}}}
\newcommand{\tsdx}{\textstyle{\frac{\rm d}{{\rm d}x}}}

\begin{document}

\title[Exact Isospectral Pairs of $\cP\cT$-Symmetric Hamiltonians]
{Exact Isospectral Pairs of $\cP\cT$-Symmetric Hamiltonians}

\author[Bender and Hook]{Carl~M~Bender${}^\ast$ and Daniel~W~Hook${}^\dag$}

\address{${}^\ast$Department of Physics, Washington University, St. Louis, MO
63130, USA \\{\footnotesize{\tt email: cmb@wustl.edu}}}

\address{${}^\dag$Theoretical Physics, Imperial College, London SW7 2AZ, UK\\
{\footnotesize{\tt email: d.hook@imperial.ac.uk}}}

\date{today}

\begin{abstract}
A technique for constructing an infinite tower of pairs of $\cP\cT$-symmetric
Hamiltonians, $\hat{H}_n$ and $\hat{K}_n$ ($n=2,\,3,\,4,\,\ldots$), that have
exactly the same eigenvalues is described and illustrated by means of three
examples ($n=2,\,3,\,4$). The eigenvalue problem for the first Hamiltonian $\hat
{H}_n$ of the pair must be posed in the complex domain, so its eigenfunctions
satisfy a complex differential equation and fulfill homogeneous boundary
conditions in Stokes' wedges in the complex plane. The eigenfunctions of the
second Hamiltonian $\hat{K}_n$ of the pair obey a real differential equation and
satisfy boundary conditions on the real axis. This equivalence constitutes a
proof that the eigenvalues of both Hamiltonians are real. Although the
eigenvalue differential equation associated with $\hat{K}_n$ is real, the
Hamiltonian $\hat{K}_n$ exhibits quantum anomalies (terms proportional to powers
of $\hbar$). These anomalies are remnants of the complex nature of the
equivalent Hamiltonian $\hat{H}_n$. For the cases $n=2,\,3,\,4$ in the classical
limit in which the anomaly terms in $\hat{K}_n$ are discarded, the pair of
Hamiltonians $H_{n,\,{\rm classical}}$ and $K_{n,\,{\rm classical}}$ have closed
classical orbits whose periods are identical. 

\end{abstract}

\submitto{\JPA}

\section{Introduction}
\label{s1}

This paper discusses the infinite class of higher-derivative Hamiltonians
\begin{equation}
\hat{H}_n=\eta\hat{p}^n-\gamma(ix)^{(n^2)}\quad(n=2,\,3,\,4,\,\ldots),
\label{eq1}
\end{equation}
where $\eta$ and $\gamma$ are arbitrary positive real parameters. These
Hamiltonians are all $\cP\cT$ symmetric; that is, they are symmetric under
combined space reflection $\cP$ and time reversal $\cT$. The literature on $\cP
\cT$ symmetry is extensive and growing rapidly; some early references
\cite{R1,R2,R3,R4} and recent review articles \cite{R5,R6,R7} provide background
for this work.

The Hamiltonians in (\ref{eq1}) are special because they are spectrally
identical to another class of Hamiltonians $\hat{K}_n$, which have entirely real
spectra. For example, for the case $n=2$ the Hamiltonian
\begin{equation}
\label{eq2}
\hat{H}_2=\frac{1}{2m}\hat{p}^2-\gamma\hat{x}^4
\end{equation}
has a real positive discrete spectrum and this spectrum is identical to the
spectrum of the conventionally Dirac Hermitian Hamiltonian
\begin{equation}
\hat{K}_2=\textstyle{\frac{1}{2m}}\hat{x}^2+4\gamma\hat{p}^4+\hbar\sqrt{
\textstyle{\frac{2\gamma}{m}}}\,\hat{p}.
\label{eq3}
\end{equation}
The spectral equivalence of these two Hamiltonians has long been known and has
been examined in many papers \cite{R8,R9,R10,R11,R12,R13,R14}.

Spectral equivalences between pairs of non-Hermitian and Hermitian Hamiltonians
have been discussed in the past \cite{R15,R16} and a number of approximately
equivalent pairs of Hamiltonians have been constructed (see, for example,
Refs.~\cite{R17,R18}). A few other exactly equivalent pairs of Hamiltonians
have been found \cite{R19,R20,R21,R22,R23,R24,R25}.

A brief demonstration that the Hamiltonians in (\ref{eq2}) and (\ref{eq3}) are
spectrally identical is given in Ref.~\cite{R11}. The prescription introduced in
Ref.~\cite{R11} makes use of several elementary transformations of differential
equations. We review this analysis here.

To begin, we construct the coordinate-space eigenvalue problem for $\hat{H}_2$
in (\ref{eq2}) by substituting $\hat{p}=-i\hbar\tsdx$. The formal eigenvalue
problem $\hat{H}_2\psi=E\psi$ then becomes the differential-equation eigenvalue
problem
\begin{equation}
\label{eq4}
-\frac{\hbar^2}{2m}\psi''(x)-\gamma x^4\psi(x)=E\psi(x).
\end{equation}
The eigenfunction $\psi(x)$ is required to satisfy homogeneous boundary
conditions inside a pair of Stokes' wedges in the complex-$x$ plane. These
wedges lie below and adjacent to the real-$x$ axis and have angular opening $\pi
/3$, as shown in Fig.~\ref{f1}.

To establish that the eigenvalues of (\ref{eq4}) are also eigenvalues of
$\hat{K}_2$ in (\ref{eq3}), we follow the simple four-step recipe introduced in
Ref.~\cite{R11}:

\begin{figure*}[b!]
\vspace{3.5in}
\includegraphics{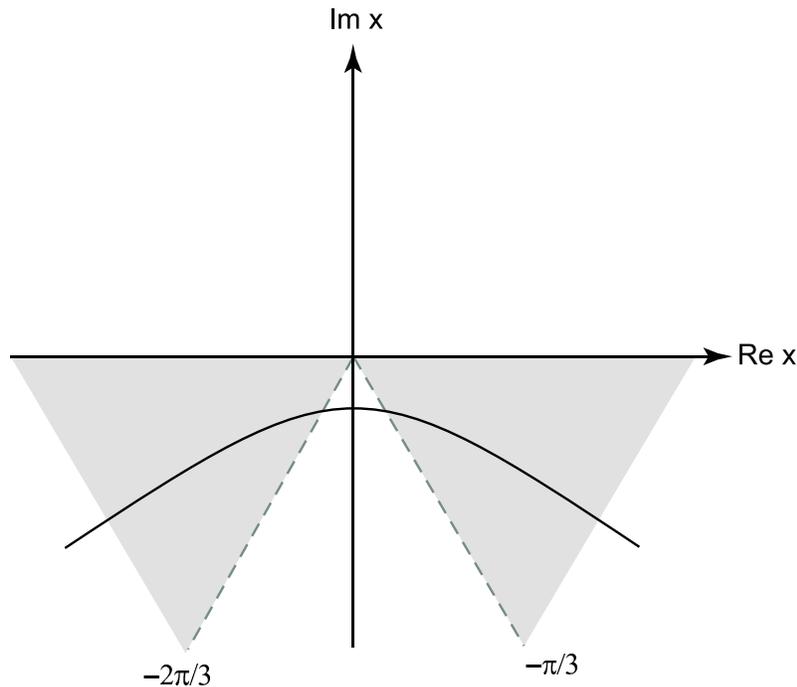}
\caption{Stokes' wedges in which the eigenfunctions $\psi(x)$ for the $-x^4$
eigenvalue problem (\ref{eq4}) are required to vanish as $|x|\to\infty$. These
wedges border on, but do not include the real axis. A possible path along which
one can solve the eigenvalue differential equation is shown. This path must
asymptote inside the Stokes' wedges.}
\label{f1}
\end{figure*}

{\bf Step 1:} Map the integration path onto the real axis by the change of
variable
\begin{equation}
x=-2i\sqrt{1+it},\quad t\in(-\infty,\infty).
\label{eq5}
\end{equation}
As $t$ runs along the real axis
from $-\infty$ to $+\infty$, $x$ runs along the complex contour that asymptotes
in the Stokes' wedges shown in Fig.~\ref{f1}. Under the change of independent
variable in (\ref{eq5}), the derivatives transform as follows:
\begin{eqnarray}
\label{eq6}
\,\,\,\tsdx&=&\sqrt{1+it}\,\tsdt,\\
\left(\tsdx\right)^2&=&(1+it)\left(\tsdt\right)^2+\half i\tsdt.
\label{eq7}
\end{eqnarray}
Thus, with the substitution $\psi(x)=\phi(t)$, (\ref{eq4}) becomes
\begin{equation}
-\textstyle{\frac{\hbar^2}{2m}}\left[(1+it)\phi''(t)+\half i\phi'(t)
\right]-16\gamma(1+it)^2\phi(t)=E\phi(t).
\label{eq8}
\end{equation}

{\bf Step 2:} Take the Fourier transform of (\ref{eq8}). The Fourier transform
of $\phi(t)$ is
\begin{equation}
f(p)=\int dt\,e^{i pt/\hbar}\phi(t),
\label{eq9}
\end{equation}
so the Fourier transform of (\ref{eq8}) is achieved by making the replacements
\begin{equation}
it\to\hbar\tsdp\quad{\rm and}\quad\tsdt\to-\frac{ip}{\hbar},
\label{eq10}
\end{equation}
and we obtain
\begin{eqnarray}
&&-16\gamma\hbar^2f''(p)+\hbar\left(\textstyle{\frac{p^2}{2m}}-32\gamma\right)
f'(p)+\left(\textstyle{\frac{p^2}{2m}}+\textstyle{\frac{3\hbar p}
{4m}}-16\gamma\right)f(p)\nonumber\\
&&\quad=Ef(p).
\label{eq11}
\end{eqnarray}

{\bf Step 3:} Introduce the new dependent variable $g(p)$ by 
\begin{equation}
f(p)=Q(p)g(p)
\label{eq12}
\end{equation}
and choose $Q(p)$ to remove the one-derivative term and thereby change
(\ref{eq11}) to Schr\"odinger form. The condition on $Q(p)$ that eliminates
terms containing $g'(p)$ is
\begin{equation}
\textstyle{\frac{Q'(p)}{Q(p)}}=\textstyle{\frac{p^2}{64m\gamma\hbar}}-
\textstyle{\frac{1}{\hbar}}.
\label{eq13}
\end{equation}
Differentiating (\ref{eq13}), we get
\begin{equation}
\textstyle{\frac{Q''(p)}{Q(p)}}=\textstyle{\frac{p}{32m\gamma\hbar}}+\left(
\textstyle{\frac{p^2}{64m\gamma\hbar}}-\textstyle{\frac{1}{\hbar}}\right)^2
\label{eq14}
\end{equation}
and substituting (\ref{eq12} -- \ref{eq14}) into (\ref{eq11}), we obtain the
Schr\"odinger equation
\begin{equation}
-16\gamma\hbar^2g''(p)+\left(\textstyle{\frac{p^4}{256m^2\gamma}}+
\textstyle{\frac{\hbar p}{4m}}\right)g(p)=Eg(p).
\label{eq15}
\end{equation}

{\bf Step 4:} Finally, rescale the independent variable $p$:
\begin{equation}
p\to\sqrt{32m\gamma}\,p.
\label{eq16}
\end{equation}
This rescaling makes (\ref{eq15}) resemble the original Schr\"odinger
equation in (\ref{eq4}):
\begin{equation}
-\textstyle{\frac{\hbar^2}{2m}}g''(p)+\left(4\gamma p^4+\hbar\sqrt{
\textstyle{\frac{2\gamma}{m}}}\,p\right)g(p)=Eg(p).
\label{eq17}
\end{equation}
The Schr\"odinger equation (\ref{eq17}) is associated with the Hamiltonian in
(\ref{eq3}). Since the differential equation (\ref{eq17}) is real and the
boundary conditions on the eigenfunctions $g(p)$ are imposed on the real axis,
it follows that the eigenvalues are real. This constitutes a rigorous proof that
the non-Hermitian Hamiltonian $\hat{H}_2$ in (\ref{eq2}) has a real spectrum.

The term in (\ref{eq3}) that is proportional to $\hbar$ is a quantum anomaly
that arises as a remnant of the complex boundary conditions on the
eigenfunctions associated with (\ref{eq4}). (These boundary conditions are
illustrated in Fig.~\ref{f1}.) In the classical limit $\hbar\to0$, we obtain the
classical Hamiltonian
\begin{equation}
K_{2,\,{\rm classical}}=\textstyle{\frac{1}{2m}}x^2+4\gamma p^4.
\label{eq18}
\end{equation}

\begin{figure*}[t!]
\vspace{3.9in}
\includegraphics{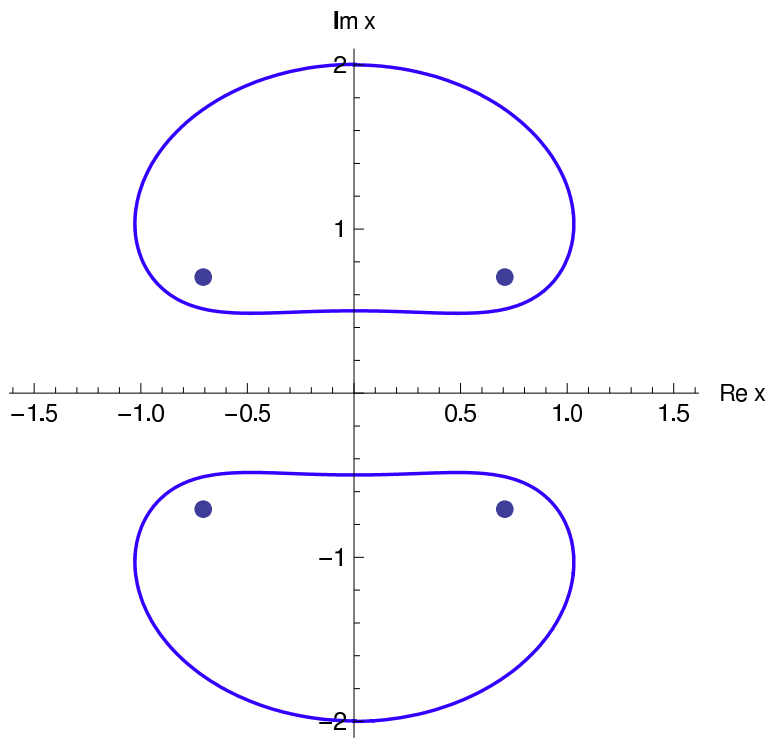}
\caption{Classical $\cP\cT$-symmetric trajectories in the complex-$x$ plane for
the Hamiltonian $H_{2,\,{\rm classical}}$ with $m=1/2$ and $\gamma=1$. For these
trajectories the energy $E=1$ and the trajectories start at $2i$ and $-2i$. The
period of the motion is $T=1.854\,074\,68\ldots\,~$.}
\label{f2}
\end{figure*}

It is an interesting but previously unnoticed fact that at the classical level
the two Hamiltonians (\ref{eq2}) and (\ref{eq18}) are equivalent. To demonstrate
this equivalence we show that the classical trajectories determined by these
Hamiltonians in complex-coordinate space have identical periods. We plot in
Fig.~\ref{f2} two classical orbits of $H_{2,\,{\rm classical}}$ for $m=1/2$,
$\gamma=1$, and energy $E=1$. Note that there are four turning points, which are
located at $x=e^{i\pi/4}$, $x=e^{3i\pi/4}$, $x=e^{5i\pi/4}$, and $x=e^{7i\pi/
4}$. A trajectory starting at $x=2i$ gives a closed orbit in the upper-half
plane, and a trajectory starting at $x=-2i$ gives a closed orbit in the
lower-half plane. Both orbits have the same period $T$, where 
\begin{equation}
T=\frac{\sqrt{2\pi}\Gamma(1/4)}{4\Gamma(3/4)}=1.854\,074\,68\ldots\,.
\label{eq19}
\end{equation}
We calculate $T$ by solving Hamilton's equations ${\dot x}=\frac{\partial H}{
\partial p}=2p$ and ${\dot p}=-\frac{\partial H}{\partial x}=4x^3$. Eliminating
$p$, we express the period $T$ as the contour integral
\begin{equation}
T=\oint\frac{dx}{2\sqrt{1+x^4}}.
\label{eq20}
\end{equation}
We then use Cauchy's theorem to distort the contour to one that goes along a ray
from the origin to a turning point, encircles the turning point, and follows the
ray back to the origin. The contour then continues along another ray out to the
other turning point, encircles this turning point, and follows that ray back to
the origin. The resulting integral becomes a standard representation for a Beta
function.

\begin{figure*}[t!]
\vspace{3.9in}
\includegraphics{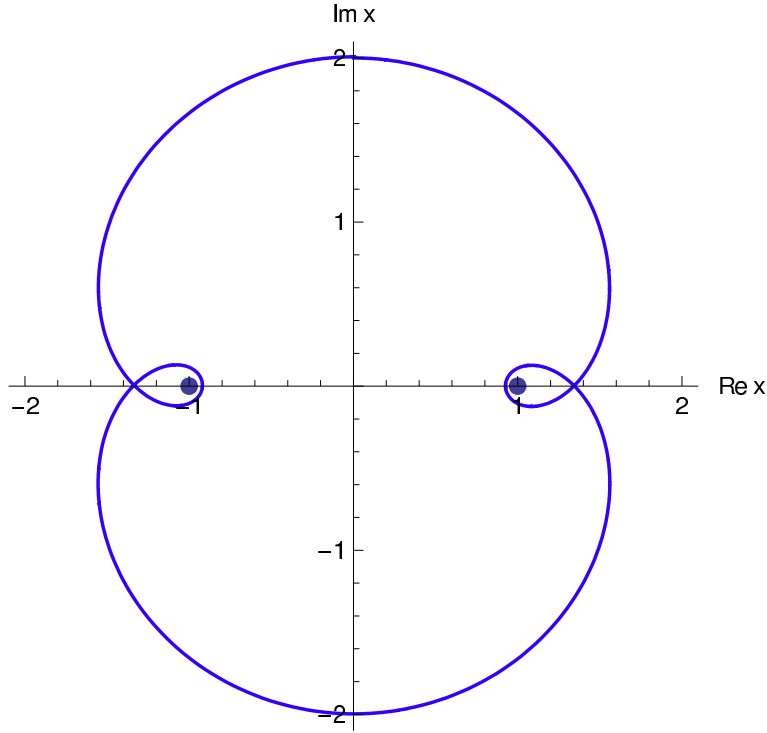}
\caption{Classical $\cP\cT$-symmetric trajectories in the complex-$x$ plane for
the $K_{2,\,{\rm classical}}$ Hamiltonian in (\ref{eq18}). The period of this
orbit is identical to the periods of the orbits shown in Fig.~\ref{f2}.}
\label{f3}
\end{figure*}

In Fig.~\ref{f3} we plot the classical orbit for a particle of energy $E=1$
described by the Hamiltonian $K_{2,\,{\rm classical}}$ in (\ref{eq18}). Note
that there are now two and not four turning points located at $x=\pm1$. The
figure shows a closed periodic classical orbit that begins at $x=2i$. The period
$T$ of this orbit is exactly that given in (\ref{eq19}).

\begin{figure*}[t!]
\vspace{1.6in}
\includegraphics{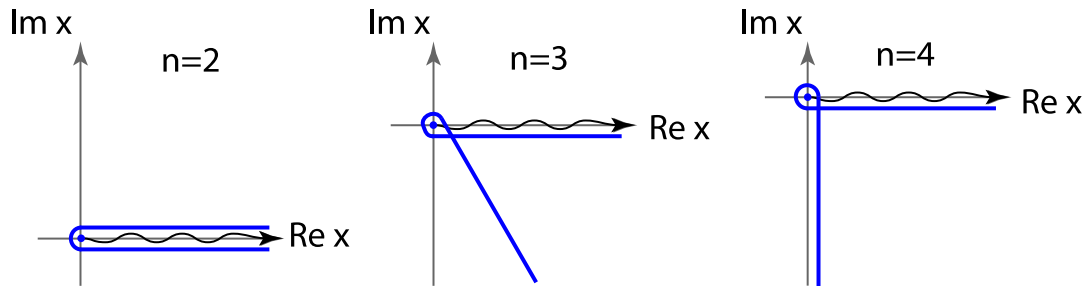}
\caption{Behavior of a classical trajectory as it approaches a turning point of 
a Hamiltonian of the form $p^n+V(x)$ for $n=2$, $n=3$, and $n=4$. When $n=2$
the trajectory executes a $180^\circ$ U-turn, when $n=3$ the trajectory makes a
$240^\circ$ turn, and when $n=4$ the trajectory makes a $270^\circ$ turn. The 
wiggly lines indicate that the turning point is a branch point of an $n$-sheeted
Riemann surface. Turning points of the $n=3$ type are shown in Figs.~\ref{f6}
and \ref{f12} and turning points of the $n=4$ type are shown in Figs.~\ref{f3}
and \ref{f8}.}
\label{f4}
\end{figure*}

Figure \ref{f3} exhibits a remarkable new topological feature that is not found
in the complex classical trajectories of Hamiltonians of the form $p^2+V(x)$,
namely, that the classical orbit makes a $270^\circ$ loop around the turning
points. There have been many studies of complex classical systems
\cite{R2,R26,R27,R28,R29,R30,R31} and in previous numerical studies of complex
trajectories the Hamiltonians that were examined were quadratic in the momentum
$p$. When the momentum term in the Hamiltonian is quadratic, the trajectory
always makes a $180^\circ$ U-turn about the turning points. In contrast, with
Hamiltonians of the form $p^3+V(x)$, the classical trajectory makes a
$240^\circ$ turn and for Hamiltonians of the form $p^4+V(x)$, the classical
trajectory makes a $270^\circ$ turn about the turning points. These behaviors
are illustrated in Fig.~\ref{f4}.

To understand the angular rotation of a classical trajectory about a turning
point, let us consider a Hamiltonian of the form $H=p^n+V(x)$. From the first of
Hamilton's equations, ${\dot x}=\frac{\partial H}{\partial p}=np^{n-1}$, we can
eliminate $p$ from the Hamiltonian and we find that a particle of energy $E$
obeys the equation ${\dot x}=n[E-V(x)]^{(n-1)/n}$. We can now see that the
turning point is also a branch point of an $n$-sheeted Riemann surface. Note
that as $n\to\infty$, a classical trajectory that approaches a turning point
will make a full $360^\circ$ turn and continue going in the same direction! (We
encounter many examples of such higher-order turning points throughout this
paper. However, we emphasize that, in general, the properties of higher-order
classical turning points are not easy to predict analytically, and require the
kind of sophisticated analysis used in catastrophe theory.)

The main objective of this paper is to show how to construct isospectral pairs
of Hamiltonians, one of which is associated with a complex eigenvalue problem
while the other is associated with a real eigenvalue problem. There have been
many attempts to find isospectral pairs of Hamiltonians of the form $\hat{p}^2+
V(\hat{x})$ by using the four-step differential-equation procedure outlined
above. However, all such attempts have proved fruitless. In this paper we
consider a wider class of Hamiltonians in which $\hat{p}^2$ is replaced by $\hat
{p}^n$, and we suggest by means of illustrative examples (rather than by
presenting a proof) that for each integer value of $n$ it is now possible to
construct such an isospectral pair of $\cP\cT$-symmetric Hamiltonians. (To
construct a proof, one would have to follow the procedure detailed in the
illustrative examples in Secs.~\ref{s3} and \ref{s4}.) The eigenfunctions for
the first member of the pair ${\hat H}_n$ satisfy boundary conditions in Stokes'
wedges in the complex plane. However, the eigenfunctions for the second member
of the pair ${\hat K}_n$ satisfy a real differential equation with homogeneous
boundary conditions given on the real axis. Therefore, our construction
constitutes a rigorous proof that entire eigenspectrum of the complex
Hamiltonian ${\hat H}_n$ is real.

The Hamiltonian $\hat{K}_n$ that is spectrally equivalent to $\hat{H}_n$ has
quantum anomaly terms containing powers of $\hbar$ up to $n-1$. If we discard
these anomaly terms, we obtain a pair of {\it classical} Hamiltonians that are
equivalent in the following sense: For each closed periodic classical trajectory
of the first Hamiltonian, it appears that there exists a closed trajectory of
the second Hamiltonian that has exactly the same period.

This paper is organized as follows: In Sec.~\ref{s2} we present some general
introductory calculations. Then, in Secs.~\ref{s3} and \ref{s4} we treat the
cases $n=3$ and $n=4$. Finally, in Sec.~\ref{s5} we make some concluding
remarks.

\section{General treatment}
\label{s2}

In order to carry out Step 1 on $\hat{H}_n$ in (\ref{eq3}) we must generalize
the mapping in (\ref{eq5}):
\begin{equation}
x=-\frac{i}{\alpha}(1+it)^\alpha,\quad t\in(-\infty,\infty),
\label{eq21}
\end{equation}
where $\alpha=1/n$. Note that this
change of variable reduces to (\ref{eq5}) when $n=2$ ($\alpha=\half$). Now, as
$t$ runs from $-\infty$ to $+\infty$ along the real axis, $x$ runs along a
complex contour that is appropriate for the Hamiltonian $\hat{H}_n$.

The transformations from $x$ derivatives to $t$ derivatives can then be written
as
\begin{eqnarray}
\,\,\,\tsdx &=& (1+it)^{1-\alpha}\tsdt,
\label{eq22}\\
\left(\tsdx\right)^2 &=& (1+it)^{2-2\alpha}\left(\tsdt\right)^2+i
(1-\alpha)(1+it)^{1-2\alpha}\tsdt,
\label{eq23}\\
\left(\tsdx\right)^3 &=&(1+it)^{3-3\alpha}\left(\tsdt\right)^3
+3i(1-\alpha)(1+it)^{2-3\alpha}\left(\tsdt\right)^2\nonumber\\
&&\qquad -(1-\alpha)(1-2\alpha)(1+it)^{1-3\alpha}\tsdt,
\label{eq24}\\
\left(\tsdx\right)^4 &=&(1+it)^{4-4\alpha}\left(\tsdt\right)^4
+6i(1-\alpha)(1+it)^{3-4\alpha}\left(\tsdt\right)^3\nonumber\\
&& \qquad -(1-\alpha)(7-11\alpha)(1+it)^{2-4\alpha}\left(\tsdt
\right)^2\nonumber\\
&& \qquad-i(1-\alpha)(1-2\alpha)(1-3\alpha)(1+it)^{1-4\alpha}\tsdt.
\label{eq25}
\end{eqnarray}
This set of equations generalizes (\ref{eq6}) and (\ref{eq7}); (\ref{eq22}) and
(\ref{eq23}) reduce to (\ref{eq6}) and (\ref{eq7}) when $\alpha=\half$. In the
next two sections we show how use these results to transform a general class of
Hamiltonians of the form (\ref{eq1}). Note that the Hamiltonian in (\ref{eq2})
corresponds to $n=2$. We begin with the simplest generalization, namely, $n=3$.

\section{Case $n=3$}
\label{s3}

In this section we consider the Hamiltonian (\ref{eq1}) for the case $n=3$:
\begin{equation}
\hat{H}_3=\eta\hat{p}^3-i\gamma \hat{x}^9.
\label{eq26}
\end{equation}
The corresponding differential-equation eigenvalue problem $\hat{H}_3\psi=E\psi$
is
\begin{equation}
\label{eq27}
i\hbar^3\eta\left(\tsdx\right)^3\psi(x)-i\gamma x^9\psi(x)=E\psi(x).
\end{equation}

\begin{figure*}[t!]
\vspace{4.2in}
\includegraphics{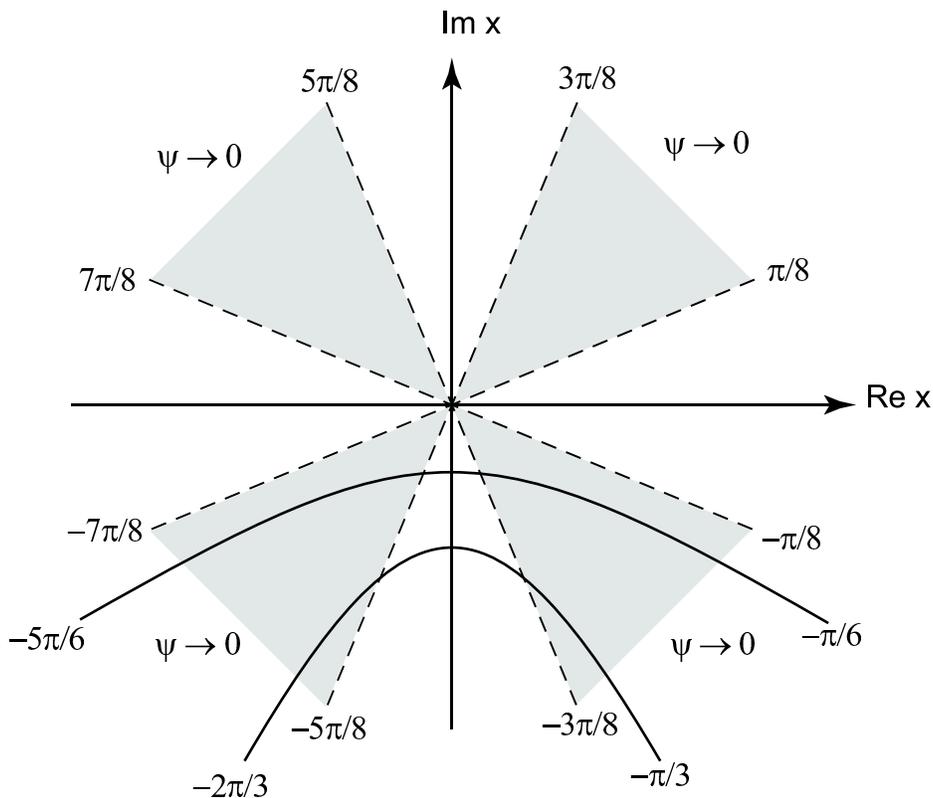}
\caption{Stokes' wedges for the solutions to the differential equation
(\ref{eq27}). The boundary conditions on the solution to this eigenvalue problem
require that $\psi(x)\to0$ as $|x|\to\infty$ with ${\rm arg}\,x$ inside the two
shaded wedges in the lower-half plane. Two integration contours inside these
wedges are shown, one of which runs from an angle of $-5\pi/6$ to an angle of
$-\pi/6$ and the other from $-2\pi/3$ to $-\pi/3$. When we set $\alpha=\frac{1}
{3}$ in (\ref{eq21}) and allow $t$ to run from $-\infty$ to $\infty$, the
variable $x$ follows the latter curve in the complex-$x$ plane.}
\label{f5}
\end{figure*}

To solve this boundary-value problem the integration path must lie in Stokes'
wedges. For the above equation these wedges do not include the real axis. The
wedges are determined by the asymptotic solutions to (\ref{eq27}), which for
large $|x|$ have the form $\exp(c\omega x^4)$, where $c=\frac{1}{4\hbar}(\gamma
/\eta)^{1/3}$ is a positive constant and $\omega$ is a cube root of unity:
$\omega^3=1$. From this asymptotic behavior we see that each of the Stokes'
wedges has angular opening $\pi/4$. We impose boundary conditions in the wedges
$-\pi/8>{\rm arg}\,x>-3\pi/8$ and $-5\pi/8>{\rm arg}\,x>-7\pi/8$. (These wedges
are the shaded regions in the lower-half plane shown in Fig.~\ref{f5}.)
Requiring that $\psi(x)$ vanish as  $|x|\to\infty$ with ${\rm arg}\,x$ in the
above two wedges eliminates two of the three possible solutions to (\ref{eq27})
and keeps only the solution whose asymptotic behavior is given by $\exp(cx^4)$.

We now follow Step 1 of the procedure explained in Sec.~\ref{s1} and make the
change of variable in (\ref{eq21}) with $\alpha=\frac{1}{3}$. From (\ref{eq24})
we then obtain the differential equation
\begin{eqnarray}
&& i\hbar^3\eta\left[(1+it)^2\left(\tsdt\right)^3+2i(1+it)\left(
\tsdt\right)^2-\textstyle{\frac{2}{9}}\tsdt\right]\phi(t)\nonumber\\
&& \qquad-3^9\gamma(1+it)^3\phi(t)=E\phi(t).
\label{eq28}
\end{eqnarray}
This differential equation is to be solved on the {\it real} axis in the $t$
variable and the solution satisfies homogeneous boundary conditions on the {\it
real} axis. Note that as $t$ runs from $-\infty$ to $\infty$, the variable $x$
runs from an angle of $-2\pi/3$ to an angle of $-\pi/3$ in the complex-$x$
plane. This integration in the complex-$x$ plane is shown in Fig.~\ref{f5}.

Following Step 2, we perform the Fourier transform in (\ref{eq9}) and use
(\ref{eq10}) to obtain
\begin{eqnarray}
&& \eta\left[-\left(1+\hbar\tsdp\right)^2p^3+2\hbar\left(1+\hbar
\tsdp\right)p^2-\textstyle{\frac{2}{9}}\hbar^2 p\right]f(p)\nonumber\\
&&\qquad-3^9\gamma\left(1+\hbar\tsdp\right)^3f(p)=Ef(p).
\label{eq29}
\end{eqnarray}

We then simplify (\ref{eq29}) by transforming it to a new differential equation
that does not have a second-derivative term. To do so we let $f(p)=Q(p)g(p)$, as
in Step 3 of Sec.~\ref{s1}. The condition on $Q(p)$ that eliminates the
second-derivative term is
\begin{equation}
\textstyle{\frac{Q'(p)}{Q(p)}}=-\textstyle{\frac{1}{\hbar}}-\textstyle{\frac{
\eta p^3}{3^{10}\gamma\hbar}},
\label{eq30}
\end{equation}
and upon differentiation we get
\begin{eqnarray}
\textstyle{\frac{Q''(p)}{Q(p)}}&=&\textstyle{\frac{1}{\hbar^2}}-\textstyle{\frac
{\eta p^2}{3^9\gamma\hbar}}+\textstyle{\frac{\eta^2p^6}{3^{20}\gamma^2\hbar^2}}
+\textstyle{\frac{2\eta p^3}{3^{10}\gamma\hbar^2}},\nonumber\\
\textstyle{\frac{Q'''(p)}{Q(p)}}&=&-\textstyle{\frac{1}{\hbar^3}}-\textstyle{
\frac{2\eta p}{3^9\gamma\hbar}}+\textstyle{\frac{\eta^2p^5}{3^{18}\gamma^2
\hbar^2}}+\textstyle{\frac{\eta p^2}{3^8\gamma\hbar^2}}
-\textstyle{\frac{\eta^2p^6}{3^{19}\gamma^2\hbar^3}}-\textstyle{\frac{
\eta p^3}{3^9\gamma\hbar^3}}-\textstyle{\frac{\eta^3p^9}{3^{30}\gamma^3\hbar^3}
}.
\label{eq31}
\end{eqnarray}
The resulting equation for $g(p)$ is 
\begin{eqnarray}
&&-3^9\gamma\hbar^3g'''(p)+\left(
\textstyle{\frac{\eta^2\hbar p^6}{3^{10}\gamma}}
-\eta\hbar^2p^2\right)g'(p)\nonumber\\
&&\qquad+\left(\textstyle{\frac{4\eta^2\hbar p^5}{3^{10}\gamma}}
-\textstyle{\frac{2\eta\hbar^2p}{9}}
-\textstyle{\frac{2\eta^3p^9}{3^{21}\gamma^2}}\right)g(p)=Eg(p).
\label{eq32}
\end{eqnarray}
This completes Step 3.

Finally, we perform the scaling
\begin{equation}
p\to\textstyle{\frac{27\gamma^{1/3}}{\eta^{1/3}}}p
\label{eq33}
\end{equation}
and obtain
\begin{eqnarray}
&&-\eta\hbar^3g'''(p)+\left(-27\eta^{2/3}\gamma^{1/3}\hbar^2p^2
+243\eta^{1/3}\gamma^{2/3}\hbar p^6\right)g'(p)\nonumber\\
&&\qquad+\left(972\eta^{1/3}\gamma^{2/3}\hbar p^5-
6\eta^{2/3}\gamma^{1/3}\hbar^2p+1458\gamma p^9\right)g(p)=Eg(p).
\label{eq34}
\end{eqnarray}
This completes Step 4 and from this result we make the {\it ansatz} $\tsdp\to
\frac{i}{\hbar}\hat{x}$ to identify the Hamiltonian that is equivalent to that
in (\ref{eq26}):
\begin{eqnarray}
&&\hat{K}_3=i\eta\hat{x}^3+i\left(-27\eta^{2/3}\gamma^{1/3}\hbar\hat{p}^2
+243\eta^{1/3}\gamma^{2/3}\hat{p}^6\right)\hat{x}\nonumber\\
&&\qquad+\left(972\eta^{1/3}\gamma^{2/3}\hbar\hat{p}^5-
6\eta^{2/3}\gamma^{1/3}\hbar^2\hat{p}+1458\gamma \hat{p}^9\right).
\label{eq35}
\end{eqnarray}
Observe that there is an $\hbar^2$ anomaly term as well as two $\hbar$ anomaly
terms. Therefore, in the classical limit $\hbar\to0$, $\hat{K}_3$ in
(\ref{eq35}) becomes
\begin{eqnarray}
K_{3,\,{\rm classical}}=i\eta x^3+243i\eta^{1/3}\gamma^{2/3}p^6x+1458\gamma p^9.
\label{eq36}
\end{eqnarray}

\begin{figure*}[t!]
\vspace{5.0in}
\includegraphics{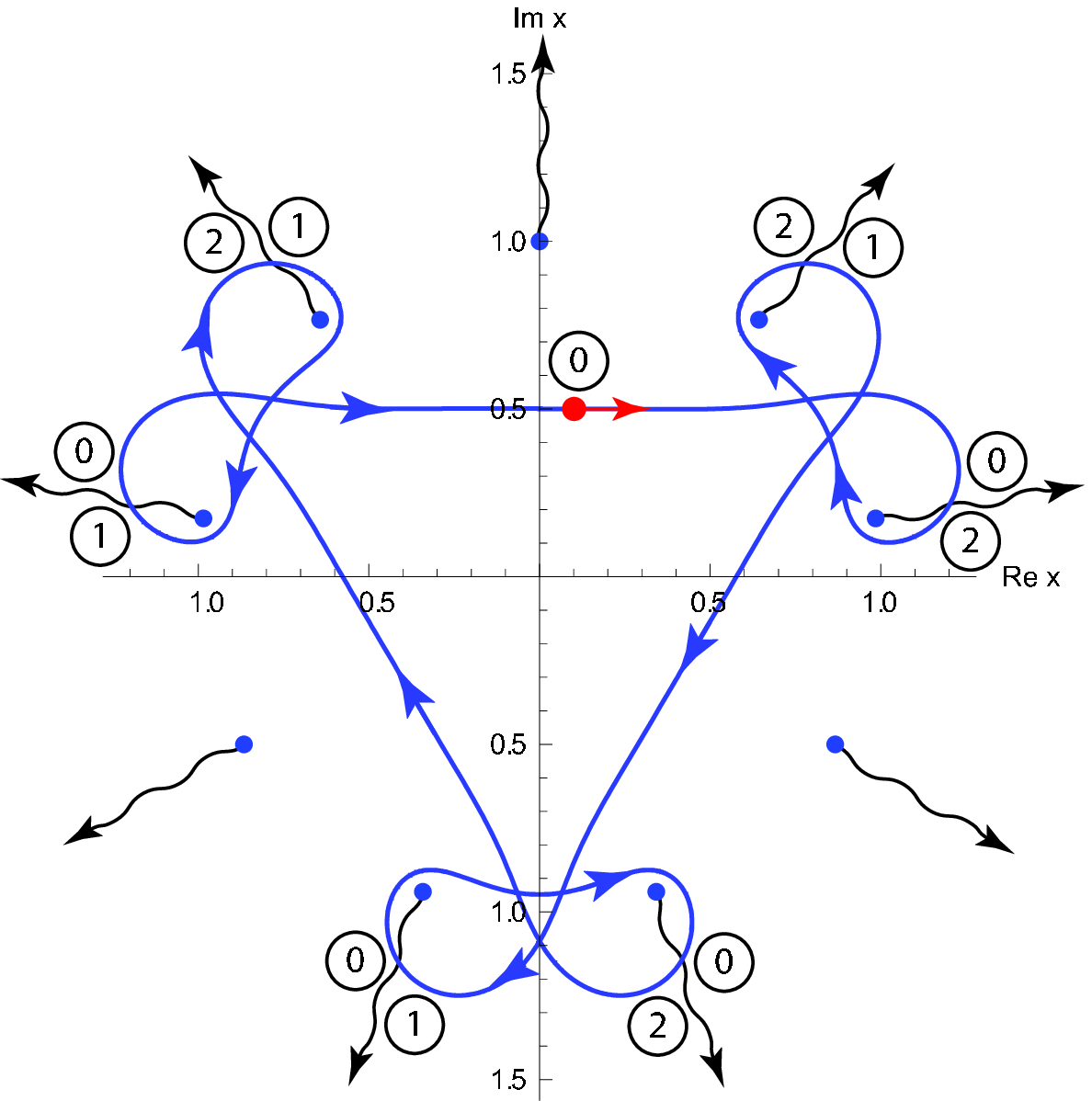}
\caption{A classical $\cP\cT$-symmetric periodic trajectory in the complex-$x$
plane for the Hamiltonian $H_{3,\,{\rm classical}}=p^3-ix^9$ for a particle of
energy $E=1$. The trajectory begins at $x=0.5\,i$ and proceeds to wind around
six of the nine turning points. It never crosses itself because each of the
turning points is a cube-root branch point, and thus the trajectory visits the
three sheets of the Riemann surface, which are numbered 0, 1, and 2. The period
of the motion is $T=3.377\,981\,999\,015\ldots$~.}
\label{f6}
\end{figure*}

We now argue that the two classical Hamiltonians, $H_{3,\,{\rm classical}}$ and
$K_{3,\,{\rm classical}}$, are equivalent by computing numerically the classical
trajectories for each Hamiltonian and verifying that trajectories having the
same energy have the same periods. In Fig.~\ref{f6} we plot a classical
trajectory for $H_{3,\,{\rm classical}}$ with $\eta=1$ and $\gamma=1$ for a
particle of energy $E=1$ starting at $x=0.5\,i$.

The trajectory in Fig.~\ref{f6} has an elaborate structure. It winds around six
of the nine turning points in the negative (clockwise) direction. Because each
of the nine turning points is also a branch point of cubic type, with each wind
the direction of the path changes by $240^\circ$, as shown in Fig.~\ref{f4}.
Thus, after the path leaves its starting point at $x=0.5\,i$, it winds in the
negative direction and crosses from sheet zero to sheet two of the Riemann
surface (which is equivalent to sheet $-1$). The number of the sheet is
indicated in a circle on Fig.~\ref{f6}. The path then heads upward and to the
left, but it does not cross itself because it is on the second sheet and no
longer on the zeroth sheet. The path then proceeds to visit each sheet of the
Riemann surface twice before returning to its starting point at $x=0.5\,i$ on
the zeroth sheet.

We can calculate the period of this orbit exactly by distorting the path shown
in Fig.~\ref{f6} into a path that begins at the origin, travels outward to a
turning point along a ray, encircles the turning point, and then returns to the
origin, and then repeats this trip five more times. However, in evaluating the
complex line integrals we must be careful to remember that the {\it phase} of
the integrand changes as the path encircles the turning point. Upon adding
together the contributions from all six turning points we obtain the exact
result $T=2[\cos(\pi/18)+\cos(7\pi/18)]\Gamma(10/9)\Gamma(1/3)/\Gamma(4/9)=
3.377\,981\,999\,015\ldots~.$

\begin{figure*}[t!]
\vspace{3.5in}
\includegraphics{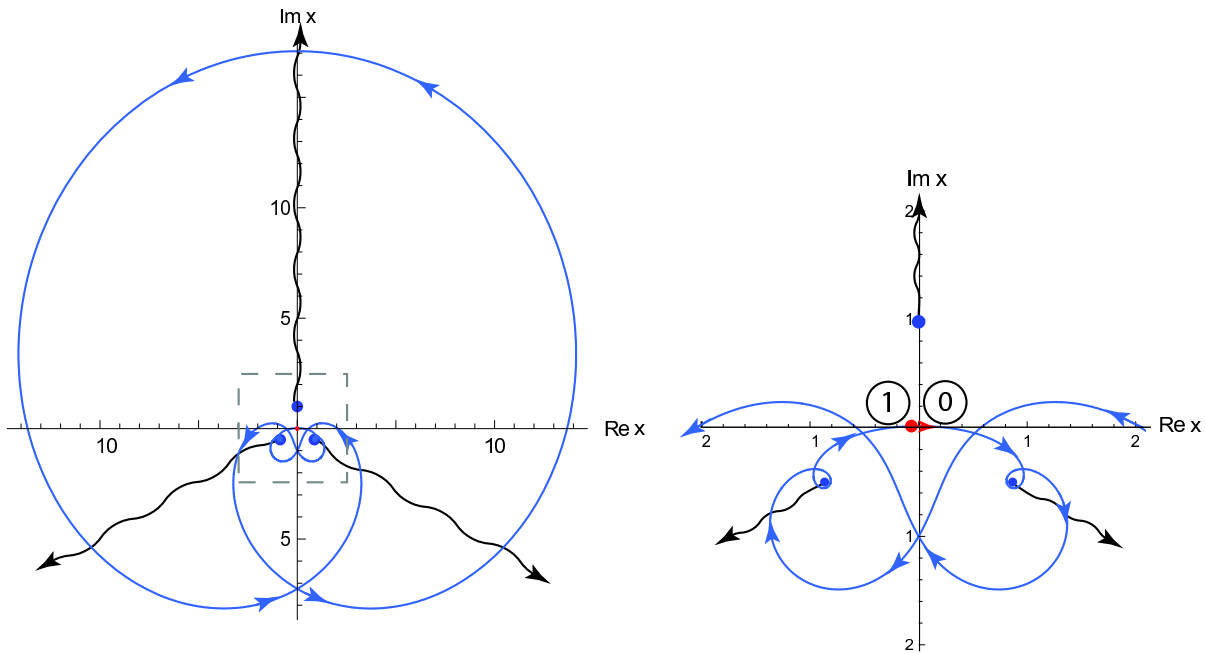}
\caption{A classical $\cP\cT$-symmetric trajectory in the complex-$x$ plane for
the Hamiltonian $K_{3,\,{\rm classical}}$ in (\ref{eq35}) with $\eta=1$ and
$\gamma=1$. The trajectory shown is that of a particle of energy $E=1$ beginning
at $x=-i$. The trajectory follows an elaborate and complicated path as it winds
around the three turning points, which are indicated by dots. The inset shows a
blown-up version of the trajectory near the origin. Note that the trajectory
never crosses itself because each of the turning points is also a cube-root
branch point, and thus the trajectory lies in a three-sheeted Riemann surface.
The period of the orbit is $T=1.125\,994\ldots$, which is exactly $1/3$ of the
period of the closed orbit shown in Fig.~\ref{f6}. It is crucial to observe that
the trajectory shown is {\it not closed}; the inset indicates that if the
trajectory is on sheet 0 just to the right of the origin, it returns to this
point on sheet 1. Thus, it must trace this path two more times before it closes.
When it finally closes, the period of its orbit agrees exactly with that of the
trajectory shown in Fig.~\ref{f6}.}
\label{f7}
\end{figure*}

In Fig.~\ref{f7} we plot the classical trajectory of a particle of energy $E=1$
governed by the Hamiltonian $K_{3,\,{\rm classical}}$ with $\eta=1$ and $\gamma=
1$. This trajectory begins at $x=0$. The time for the particle to follow the
path shown in this figure is $T=1.125\,994\ldots$, which is precisely $1/3$ of
the period of the closed orbit shown in Fig.~\ref{f6}. We emphasize that the
trajectory of this orbit is {\it not yet closed}. Figure \ref{f7} shows that the
particle must trace this path three times before the orbit closes, and the time
required to do this is exactly the period of the orbit shown in Fig.~\ref{f6}.

One might think (wrongly!) that there should be a total of {\it six} rather than
three turning points on Fig.~\ref{f7}. We determine the positions of the turning
points from the first of Hamilton's equations
\begin{equation}
\dot{x}=\frac{\partial}{\partial p}K_{3,\,{\rm classical}}=3^6 2ip^5x+3^8 2 p^8.
\label{eq37}
\end{equation}
The turning points are the points where $\dot{x}=0$, and this seems to occur
when $p=0$ and when $ix+9p^3=0$. However, if we then substitute each of these
conditions into the equation $K_{3,\,{\rm classical}}=1$, we find that only the
first of these conditions is consistent, so we learn that there are three
turning points situated at the three roots of $-i$. The second condition is
inconsistent.

\section{Case $n=4$}
\label{s4}

In this section we consider the Hamiltonian
\begin{equation}
\hat{H}_4=\eta\hat{p}^4-\gamma\hat{x}^{16}.
\label{eq38}
\end{equation}
The substitution $\hat{p}=-i\hbar\tsdx$ converts the formal eigenvalue
problem $\hat{H}_4\psi=E\psi$ into the differential-equation eigenvalue problem
\begin{equation}
\eta\hbar^4\left(\tsdx\right)^4\psi(x)-\gamma x^{16}\psi(x)=E\psi(x).
\label{eq39}
\end{equation}
The asymptotic behavior of the eigenfunctions $\psi(x)$ has the form
$\exp(icx^5/5)$, where $c=\frac{1}{5\hbar}(\gamma/\eta)^{1/4}$ is a
positive constant. Thus, the eigenfunctions vanish exponentially fast in a
pair of $\cP\cT$-symmetric wedges in the lower-half complex-$x$ plane
centered about $-3\pi/10$ and $-7\pi/10$.

{\bf Step 1:} We map this differential-equation eigenvalue problem onto the real
axis by taking $\alpha=\frac{1}{4}$ in (\ref{eq21}) so that $x=(1-it)^{1/4}$.
The mapped eigenvalue problem is
\begin{eqnarray}
&&\eta\hbar^4\left[(1+it)^3\left(\tsdt\right)^4+\textstyle{\frac{9}
{2}}i(1+it)^2\left(\tsdt\right)^3-\textstyle{\frac{51}{16}}(1+it)\left(\tsdt
\right)^2-\textstyle{\frac{3}{32}}i\tsdt\right]\phi(t)\nonumber\\
&&\qquad -4^{16}\gamma(1+it)^4\phi(t)=E\phi(t).
\label{eq40}
\end{eqnarray}

{\bf Step 2:} Taking the Fourier Transform of (\ref{eq40}), we get
\begin{eqnarray}
&&\eta\hbar^4\left[\left(1+\hbar\tsdp\right)^3\textstyle{\frac{p^4}{\hbar^4
}}-\textstyle{\frac{9}{2}}\left(1+\hbar\tsdp\right)^2\textstyle{\frac{p^3}{
\hbar^3}}+\textstyle{\frac{51}{16}}\left(1+\hbar\tsdp\right)\textstyle{\frac{
p^2}{\hbar^2}}-\textstyle{\frac{3}{32}}\textstyle{\frac{p}{\hbar}}\right]f(p)
\nonumber\\
&&\qquad -4^{16}\gamma\left(1+\hbar\tsdp\right)^4f(p)=Ef(p).
\label{eq41}
\end{eqnarray}

{\bf Step 3:} We transform the dependent variable using $f(p)=Q(p)g(p)$ and
choose $Q(p)$ to eliminate the third-derivative term $g'''(p)$. The function
$Q(p)$ that does this is
\begin{equation}
Q(p)=\exp\left(-\textstyle{\frac{p}{\hbar}}+\textstyle{\frac{\eta p^5}{4^{17}5
\gamma\hbar}}\right).
\label{eq42}
\end{equation}
The result in (\ref{eq42}) simplifies (\ref{eq41}) to 
\begin{eqnarray}
&&-4^{16}\hbar^4\gamma g''''(p)+\textstyle{\frac{3p^{16}\eta^4}{4^{52}\gamma^3}}
g(p)+\textstyle{\frac{\hbar\eta^3}{4^{35}\gamma^2}}\left[8p^{12}g'(p)+54p^{11}
g(p)\right]\nonumber\\
&&\quad+\textstyle{\frac{\hbar^2\eta^2}{4^{19}\gamma}}\left[24p^8g''(p)+240p^7
g'(p)+483p^6g(p)\right]\nonumber\\
&&\quad-\textstyle{\frac{\hbar^3\eta}{32}}\left[-48p^3g''(p)-6p^2g'(p)
+87pg(p)\right]=Eg(p).
\label{eq43}
\end{eqnarray}

{\bf Step 4:} Next, we perform the scaling transformation
\begin{equation}
p\to\textstyle{\frac{4^4\gamma^{1/4}}{\eta^{1/4}}}p
\label{eq44}
\end{equation}
and obtain the {\it real} eigenvalue problem
\begin{eqnarray}
&&-\eta\hbar^4g''''(p)+4^{12}3\gamma p^{16}g(p)+\hbar4^9\gamma^{3/4}\eta^{1/4}
\left[8p^{12}g'(p)+54p^{11}g(p)\right]\nonumber\\
&&\quad+\hbar^2 4^5\sqrt{\gamma\eta}\left[24p^8g''(p)+240p^7g'(p)+483p^6g(p)
\right]\nonumber\\
&&\quad-8\hbar^3\gamma^{1/4}\eta^{3/4}\left[-48p^3g''(p)-6p^2g'(p)+87pg(p)
\right]=Eg(p),
\label{eq45}
\end{eqnarray}
whose boundary conditions are posed on the real axis. Finally, we identify the
Hamiltonian that gives rise to this eigenvalue problem by substituting $\tsdp\to
\frac{i}{\hbar}\hat{x}$:
\begin{eqnarray}
\hat{K}_4&=&-\eta\hat{x}^4+4^{12}3\gamma \hat{p}^{16}
+4^9\gamma^{3/4}\eta^{1/4}\left(8i\hat{p}^{12}\hat{x}
+54\hbar\hat{p}^{11}\right)\nonumber\\
&&\quad+4^5\sqrt{\gamma\eta}\left(-24\hat{p}^8\hat{x}^2+240\hbar i\hat{p}^7
\hat{x}+483\hbar^2\hat{p}^6\right)\nonumber\\
&&\quad-8\gamma^{1/4}\eta^{3/4}\left(48\hbar \hat{p}^3\hat{x}^2-6i\hbar^2
\hat{p}^2\hat{x}+87\hbar^3\hat{p}\right).
\label{eq46}
\end{eqnarray}
We have thus shown that the eigenvalues of $\hat{K}_4$ are real and are
identical to the eigenvalues of $\hat{H}_4$.

Notice that $\hat{K}_4$ has first-, second-, and third-order anomaly terms. In
the classical limit $\hbar\to0$ we obtain
\begin{equation}
K_{4,\,{\rm classical}}=-2^{13}3\sqrt{\gamma\eta}\,p^8x^2
-\eta x^4+2^{21}i\gamma^{3/4}\eta^{1/4}p^{12}x+2^{24}3\gamma p^{16}.
\label{eq47}
\end{equation}
We will now demonstrate that the two classical Hamiltonians, $H_{4,\,{\rm
classical}}$ and $K_{4,\,{\rm classical}}$, are equivalent by showing that the
classical periods of the motion are identical.

\begin{figure*}[t!]
\vspace{4.7in}
\includegraphics{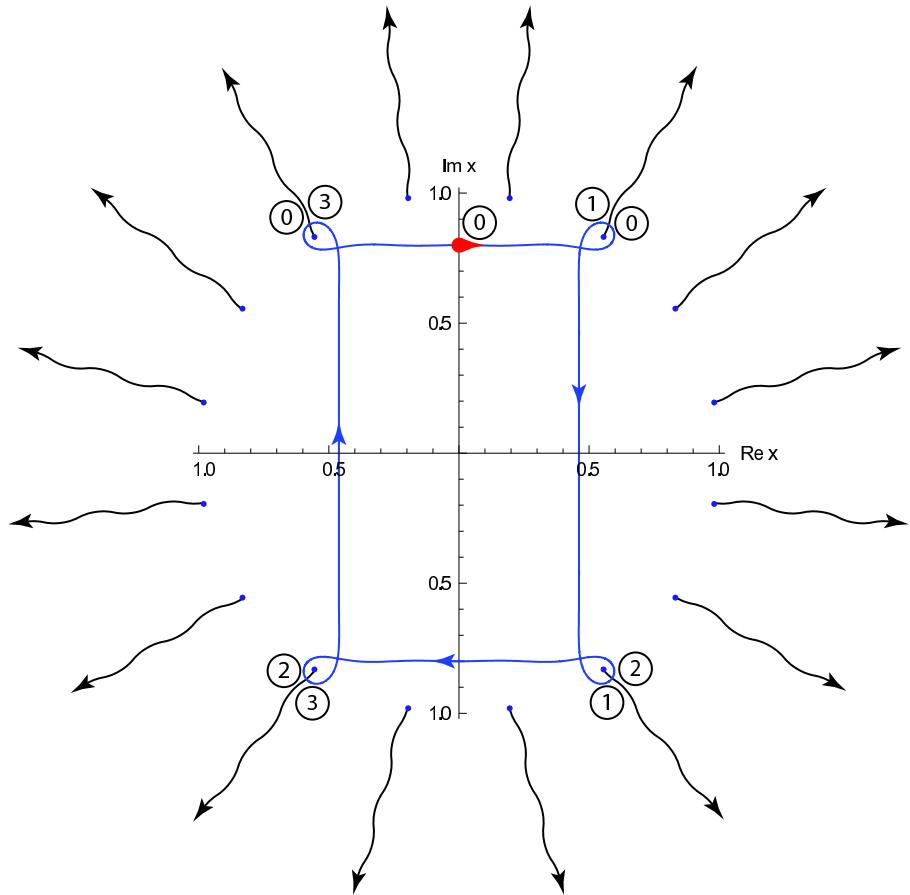}
\caption{A classical $\cP\cT$-symmetric periodic trajectory in the complex-$x$
plane for the Hamiltonian $H_{4,\,{\rm classical}}=p^4-x^{16}$ for a particle of
energy $E=1$. The trajectory begins at $x=0.8i$ and proceeds to wind around four
of the 16 turning points. It never crosses itself because each of the turning
points is a fourth-root branch point, and thus the trajectory visits the four 
sheets of the Riemann surface, which are numbered 0, 1, 2, and 3. The period of
the motion is $T=1.697\,723\,019\,533\ldots$~.}
\label{f8}
\end{figure*}

Let us first examine the complex particle motion due to $H_{4,\,{\rm
classical}}$ for the case $\eta=1$ and $\gamma=1$. In Fig.~\ref{f8} we plot a
closed orbit of a particle of energy $E=1$ that begins at $0.8i$. The orbit
takes the approximate shape of a rectangle and makes $270^\circ$ turns at
the four corners. The period of this orbit is $T=1.697\,723\,019\,533\ldots$~.
We can calculate this period exactly by distorting it into rays that go from
the origin out to each of the turning points and back to the origin, taking
great care to keep track of the phase of the integrand. The exact formula
for the period is $T=[\cos(3\pi/16)+\cos(5\pi/16)]\Gamma(1/4)\Gamma(17/16)/
\Gamma(5/16)$.

\begin{figure*}[t!]
\vspace{3.8in}
\includegraphics{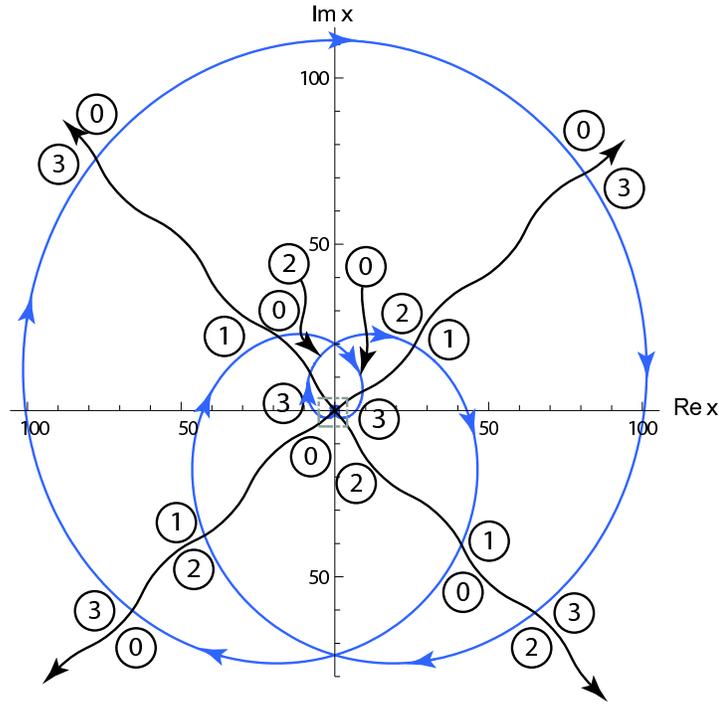}
\caption{A classical trajectory in the complex-$x$ plane for the Hamiltonian
$K_{4,\,{\rm classical}}$ in (\ref{eq47}) with $\eta=1$ and $\gamma=1$. The
trajectory represents a particle of energy $E=1$ that begins at $x=e^{-i\pi/4}$.
The trajectory winds around the turning points in a complicated fashion (see
nested insets in Figs.~\ref{f10} and \ref{f11}) but it never crosses itself
because it is on a multisheeted Riemann surface with each of the turning points
being a fourth-root branch point. The trajectory returns to its starting point
in time $T=0.848\,86$, but it is not a closed trajectory; the trajectory begins
on sheet 0 and returns on sheet two of the Riemann surface. To complete its
periodic motion the trajectory must follow the same path one more time. The time
needed to complete this double loop is exactly equal to the period of the orbit
shown in Fig.~\ref{f8}.}
\label{f9}
\end{figure*}

\begin{figure*}[t!]
\vspace{4.0in}
\includegraphics{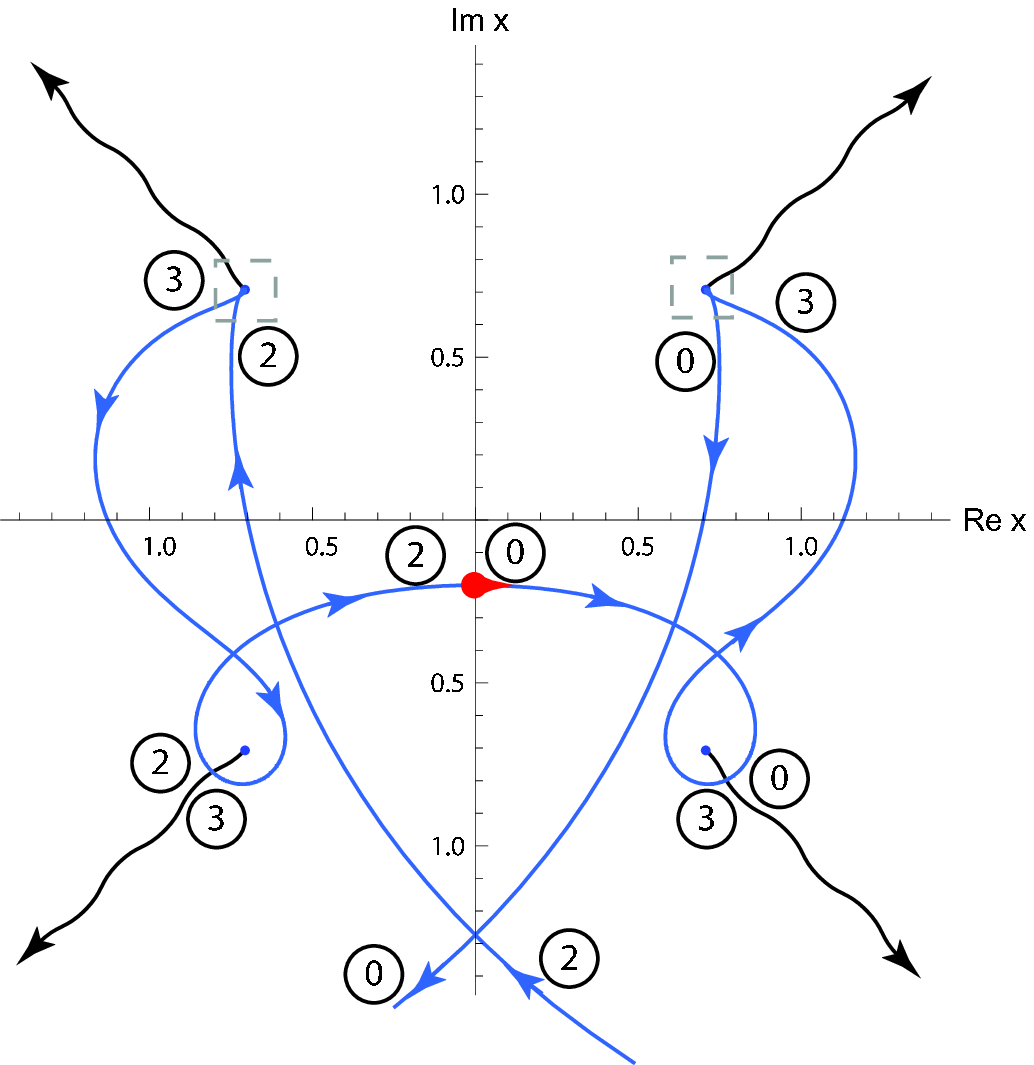}
\caption{A blown-up view of the trajectory shown in Fig.~\ref{f9} (see inset in
Fig.~\ref{f9}). The four fourth-root turning points are visible in this figure.
However, the trajectory has a complicated structure near the upper two turning
points, and Fig.~\ref{f11} shows a detail of the small region surrounding the
upper-right turning point.}
\label{f10}
\end{figure*}

Next, we examine the classical trajectories of $K_{4,\,{\rm classical}}$ for the
case $\eta=1$ and $\gamma=1$. In Fig.~\ref{f9} we plot the classical trajectory
for a particle of $E=1$ that begins at $x=-i$. There are four turning points of
fourth-root type. However, in this figure it is not possible to see the detailed
structure of this trajectory near the origin. Therefore, in Fig.~\ref{f10} we
plot a blow-up of the square region in Fig.~\ref{f9} near the origin.

\begin{figure*}[t!]
\vspace{3.3in}
\includegraphics{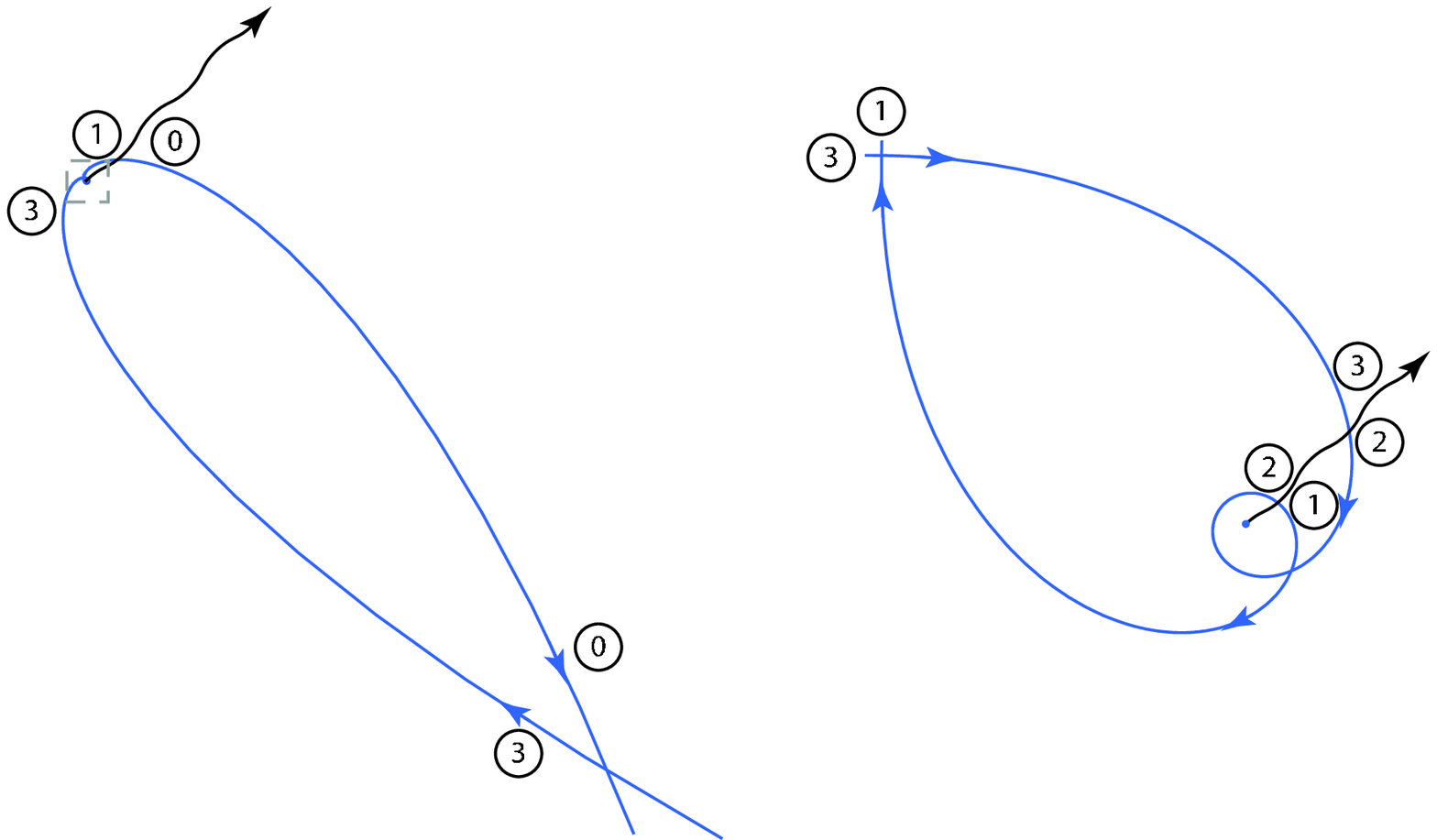}
\caption{Left: A blown-up view of a portion of Fig.~\ref{f10} (see inset in
Fig.~\ref{f10}). Right: A blown-up view of the inset on the left side.}
\label{f11}
\end{figure*}

In Fig.~\ref{f10} we can now see the four turning points and a rather
complicated trajectory. The regions surrounding the upper pair of turning points
have a highly detailed structure, and thus the square region surrounding the
upper-right turning point is blown up again in Fig.~\ref{f11} (left side). The
vicinity of the turning point contains additional structure, and this region
must be blown up still more in Fig.~\ref{f11} (right side).

\section{Summary}
\label{s5}

We have shown by using a number of examples that it is possible to construct an
infinite tower of pairs of isospectral Hamiltonians $\hat{H}_n$ and $\hat{K}_n$,
for which the first member of the pair is a complex $\cP\cT$-symmetric
Hamiltonian. The differential-equation eigenvalue problem for the second
Hamiltonian is entirely real, and therefore the eigenvalues of both Hamiltonians
are real. The second member of the pair has quantum anomalies of order
$\hbar^{n-1}$. We have also shown that at the classical level, where the
anomaly terms in $\hat{K}_n$ are discarded, the Hamiltonians are equivalent by
demonstrating that they have closed orbits of the same period.

\begin{figure*}[t!]
\vspace{5.0in}
\includegraphics{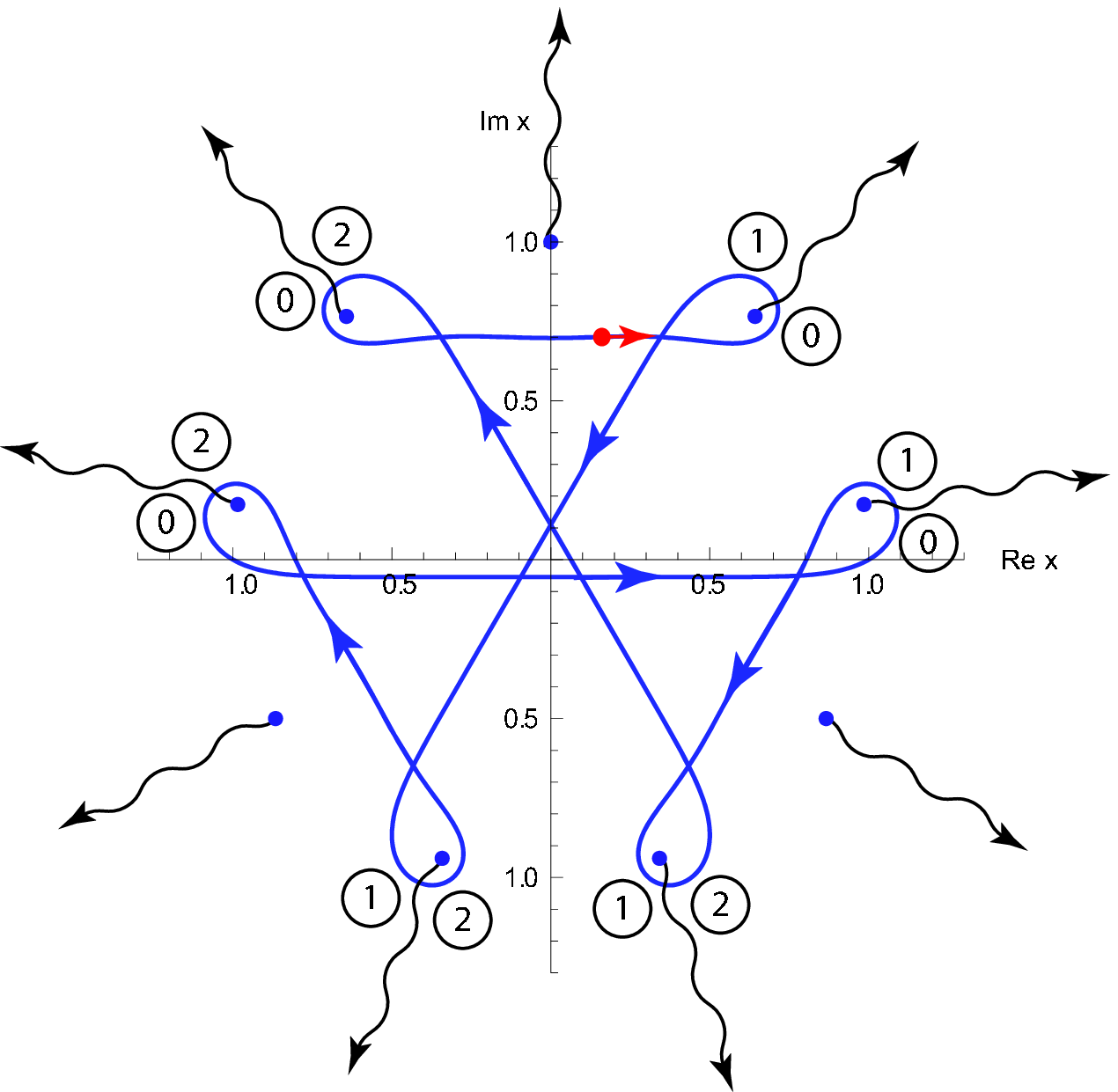}
\caption{A classical $\cP\cT$-symmetric periodic trajectory in the complex-$x$
plane for the Hamiltonian $H_{3,\,{\rm classical}}=p^3-ix^9$ for a particle of
energy $E=1$. The trajectory begins at $x=1$ and proceeds to wind around six of
the nine turning points. It never crosses itself because each of the turning
points is a cube-root branch point, and thus the trajectory visits the three
sheets of a Riemann surface, which are numbered 0, 1, and 2. The period of the 
motion is $T=4.143\,708\,353\ldots$~.}
\label{f12}
\end{figure*}

Many unsolved problems remain. For example, we do not know if for {\it every}
classical orbit of the Hamiltonians $H_{n,\,{\rm classical}}$ there is a
corresponding classical orbit of the Hamiltonians $K_{n,\,{\rm classical}}$ of
exactly the same period. Because complex coordinate space is a Riemann surface
having an elaborate sheet structure, it is possible to find various classical
orbits having many different periods. For example, for the classical Hamiltonian
$H_{3,\,{\rm classical}}$ with $\eta=1$ and $\gamma=1$, we find that a particle
of energy $E=1$ that starts at $x=1$ has a closed trajectory of period
$T=6[\cos(\pi/18)+\cos(5\pi/18)]\Gamma(10/9)\Gamma(4/3)/\Gamma(4/9)=
4.143\,708\,353\ldots$~. This orbit is shown in Fig.~\ref{f12}.

We calculate the period of this orbit exactly by distorting the path shown
in Fig.~\ref{f12} into a path that begins at the origin, travels outward to a
turning point along a ray, encircles the turning point, and then returns to the
origin, and then repeats this trip five more times. In the complex line
integrals the {\it phase} for each path advances as the path encircles the
turning point. Upon adding together the contributions from all six turning
points we obtain an analytic expression for the period of the motion. Although
we are convinced that one exists, we have not yet been able to find a
corresponding orbit of the Hamiltonian $K_{3,\,{\rm classical}}$.

\vspace{0.5cm}
\footnotesize
\noindent
CMB is supported by a grant from the U.S. Department of Energy.

\end{document}